\documentclass{article}%
\usepackage{amsmath}
\usepackage{amsfonts}
\usepackage{amssymb}
\usepackage{graphicx}%
\providecommand{\U}[1]{\protect\rule{.1in}{.1in}}
\begin{document}

\title{\vspace{-0.5in}\hfill{\normalsize ANL-HEP-PR-08-35}\bigskip\bigskip\\\vspace{-0.5in}\hfill{\normalsize UMTG - 7\bigskip}\\Ternary Virasoro - Witt Algebra}
\author{Thomas L Curtright,$^{\text{\S ,\#}}$ \ David B Fairlie,$^{\text{\dag,\#}}%
$\medskip\ and Cosmas K Zachos$^{\text{\P ,\#}}$\\$^{\text{\S }}${\small Department of Physics, University of Miami, Coral
Gables, Florida 33124, USA}\\$^{\text{\dag}}${\small Department of Mathematical Sciences, Durham
University, Durham, DH1 3LE, UK}\\$^{\text{\P }}${\small High Energy Physics Division, Argonne National
Laboratory, Argonne, IL 60439-4815, USA} }
\date{}
\maketitle

\begin{abstract}
A 3-bracket variant of the Virasoro-Witt algebra is constructed through the
use of su(1,1) enveloping algebra techniques. \ The Leibniz rules for
3-brackets acting on other 3-brackets in the algebra are discussed and
verified in various situations.

\vfill

\underline
{$\ \ \ \ \ \ \ \ \ \ \ \ \ \ \ \ \ \ \ \ \ \ \ \ \ \ \ \ \ \ \ \ \ \ \ \ \ \ \ \ \ \ \ \ \ \ \ \ \ \ \ \ \ \ \ \ \ \ \ \ \ \ \ \ \ \ \ \ \ \ \ \ \ $%
}

$^{\text{\#}}${\footnotesize curtright@miami.edu, david.fairlie@durham.ac.uk,
zachos@anl.gov}

\bigskip

\end{abstract}

\section{Introduction}

Recently there has been some progress in constructing a $2+1$-dimensional
local quantum field theory with $SO(8)$ superconformal symmetry
\cite{B,S,G,Gustavsson,H}. \ This is a useful stepping-stone to obtain a
world-volume Lagrangian description for coincident M2-branes. \ Crucial to the
construction is the use of 3-algebras (a term originally coined by Filippov
\cite{F}, following up on earlier work of Nambu \cite{N}) which are built
around antisymmetrized products of three operators. \ It now seems possible to
incorporate an arbitrary Lie algebra into a 3-algebra through the use of a
small number of auxiliary charges.

While it may be sufficient to restrict attention to finite Lie and 3-algebras
for model-building purposes, e.g. in M-theory, nevertheless it is
intrinsically interesting to also consider infinite dimensional 3-algebras.
\ Previous studies \cite{H}\ have already considered the infinite 3-algebra
that follows from the classical Nambu 3-bracket, on a 3-torus, in the form%
\begin{equation}
\left[  E_{a},E_{b},E_{c}\right]  =a\cdot\left(  b\times c\right)
E_{a+b+c}\ . \label{ClassicalNambu3BracketAlgebra}%
\end{equation}
The indices here are 3-vectors, with $\cdot$\ and $\times$\ the usual dot and
cross products of those 3-vectors. \ By virtue of some simple identities for
the volumes of parallelepipeds, this satisfies Filippov's condition\ (the
FI,\ or so-called \textquotedblleft fundamental identity\textquotedblright) --
a property he required to warrant the 3-algebra designation.
\begin{equation}
\left[  E_{x},E_{y},\left[  E_{a},E_{b},E_{c}\right]  \right]  =\left[
\left[  E_{x},E_{y},E_{a}\right]  ,E_{b},E_{c}\right]  +\left[  E_{a},\left[
E_{x},E_{y},E_{b}\right]  ,E_{c}\right]  +\left[  E_{a},E_{b},\left[
E_{x},E_{y},E_{c}\right]  \right]  \ .
\label{ClassicalNambu3BracketAlgebraFIs}%
\end{equation}
This is easily remembered as just a Leibniz rule for 3-brackets acting on
other 3-brackets. \ (However, this does \emph{not} necessarily mean $\left[
E_{a},E_{b},AB\right]  $ is the same as $\left[  E_{a},E_{b},A\right]
B+A\left[  E_{a},E_{b},B\right]  $ for arbitrary $A$ and $B$. \ While these
are the same for the \emph{classical} Nambu bracket, in general these two
expressions differ for an operator algebra.)

A 3-algebraic variant of the $su\left(  2\right)  $ Kac-Moody algebra has also
been studied \cite{KM}. \ Moreover, it is straightforward to embed an infinite
Onsager 3-algebra in this framework, just as it is to embed the usual infinite
Onsager Lie algebra \cite{Onsager} (also see \cite{FZ}) into that for
$su\left(  2\right)  $. \ Nonetheless, it is not completely clear from these
previous examples how infinite conformal 3-algebras should be constructed.
\ We consider this problem here. \ We construct an infinite 3-algebra variant
of the centerless Virasoro (i.e. Witt) algebra. \ We leave the inclusion of
central charges as well as supersymmetric extensions of our results for future work.

\section{Three-brackets from Enveloping Algebras}

The \textquotedblleft BL algebra\textquotedblright\ presented by Bagger and
Lambert \cite{B} is actually just a minor modification of Nambu's original
results\ on 3-brackets for the su(2) angular momentum algebra, $\left[
L_{x},L_{y}\right]  =iL_{z}$, etc. \ Of course, Nambu's work motivated and
inspired both Filippov (12 years later) and Takhtajan (20 years later)
\cite{T}, as well as more recent studies (see \cite{C} and references
therein). \ 

After introducing the 3-bracket, Nambu noted that the $su(2)$ Casimir followed
immediately from it (see Eqn(49) in \cite{N}).
\begin{equation}
\left[  L_{x},L_{y},L_{z}\right]  \equiv L_{x}\left[  L_{y},L_{z}\right]
+L_{y}\left[  L_{z},L_{x}\right]  +L_{z}\left[  L_{x},L_{y}\right]  =i\left(
L_{x}^{2}+L_{y}^{2}+L_{z}^{2}\right)  =iL^{2}\ .
\end{equation}
From this it follows that the BL$\ $algebra can be easily found in the
enveloping algebra for $su\left(  2\right)  $. \ If we rescale the angular
momenta by a fourth root of the Casimir,%
\begin{equation}
Q_{x}=\frac{L_{x}}{\sqrt[4]{L^{2}}}\ ,\ \ \ Q_{y}=\frac{L_{y}}{\sqrt[4]{L^{2}%
}}\ ,\ \ \ Q_{z}=\frac{L_{z}}{\sqrt[4]{L^{2}}}\ , \label{NambuSU2}%
\end{equation}
as well as define a fourth charge as that fourth root,%
\begin{equation}
Q_{t}=\sqrt[4]{L^{2}}\ , \label{NambuA4}%
\end{equation}
then we obtain $\left[  Q_{t},Q_{x}\right]  =0$ etc. \ Hence%
\begin{equation}
\left[  Q_{x},Q_{y},Q_{z}\right]  =iQ_{t}\ ,\ \ \ \left[  Q_{t},Q_{x}%
,Q_{y}\right]  =iQ_{z}\ ,\ \ \ \left[  Q_{t},Q_{y},Q_{z}\right]
=iQ_{x}\ ,\ \ \ \left[  Q_{t},Q_{z},Q_{x}\right]  =iQ_{y}\ .
\label{A4OurPhases}%
\end{equation}
So, any matrix rep of the angular momentum algebra yields a matrix rep of this
3-bracket algebra. \ This algebra can be summarized as
\begin{equation}
\left[  Q_{a},Q_{b},Q_{c}\right]  =i\varepsilon_{abc}^{\ \ \ \ d}~Q_{d}\ ,
\end{equation}
where $\varepsilon_{xyzt}=+1$ with a $\left\lceil -1,-1,-1,+1\right\rfloor $
Lorentz signature used to raise indices. \ In this form it is easily verified
that the FI conditions are satisfied: \ $\left[  Q_{d},Q_{e},\left[
Q_{a},Q_{b},Q_{c}\right]  \right]  =\left[  \left[  Q_{d},Q_{e},Q_{a}\right]
,Q_{b},Q_{c}\right]  +\left[  Q_{a},\left[  Q_{d},Q_{e},Q_{b}\right]
,Q_{c}\right]  +\left[  Q_{a},Q_{b}\left[  Q_{d},Q_{e},Q_{c}\right]  \right]
$. \ (Again, this does \emph{not} mean $\left[  Q_{d},Q_{e},AB\right]  $ is
the same as $\left[  Q_{d},Q_{e},A\right]  B+A\left[  Q_{d},Q_{e},B\right]  $
for arbitrary $A$ and $B$.)

Let us now pursue such enveloping algebra techniques to consider infinite
conformal algebras as realized nonlinearly with $su\left(  1,1\right)  $
generators. \ This gives the \emph{centerless}\ version of the usual Virasoro
algebra,%
\begin{equation}
\left[  L_{n},L_{m}\right]  =\left(  n-m\right)  L_{n+m}\ .
\label{VirasoroWitt}%
\end{equation}
An explicit realization is \cite{Fairlie}%
\begin{align}
L_{n}  &  =\left(  L_{0}+n\beta\right)  \frac{\Gamma\left(  L_{0}%
+\beta\right)  }{\Gamma\left(  L_{0}+\beta+n\right)  }\left(  L_{1}\right)
^{n}\ ,\nonumber\\
L_{-n}  &  =\left(  L_{0}-n\beta\right)  \frac{\Gamma\left(  L_{0}%
+1-\beta-n\right)  }{\Gamma\left(  L_{0}+1-\beta\right)  }\left(
L_{-1}\right)  ^{n}\ ,
\end{align}
where the $su\left(  1,1\right)  $ algebra and Casimir are given by%
\begin{align}
\left[  L_{+1},L_{-1}\right]   &  =2L_{0}\ ,\ \ \ \left[  L_{\pm1}%
,L_{0}\right]  =\pm L_{\pm1}\ ,\nonumber\\
\beta\left(  1-\beta\right)   &  \equiv C=L_{+1}L_{-1}-\left(  L_{0}+1\right)
L_{0}\ . \label{su(1,1)}%
\end{align}
From this construction, we compute 3-brackets and abstract from them a
3-algebra. \ An interesting question is: Does the algebra so obtained satisfy
the 3-bracket Leibniz rule when acting on other 3-brackets? \ 

First we compute the 3-bracket for charges with positive indices.%
\begin{align}
\left[  L_{k},L_{m},L_{n}\right]   &  =\left(  m-n\right)  L_{k}%
L_{m+n}+\left(  n-k\right)  L_{m}L_{n+k}+\left(  k-m\right)  L_{n}%
L_{k+m}\nonumber\\
&  =C~\frac{\left(  k-m\right)  \left(  m-n\right)  \left(  k-n\right)
}{L_{0}+\left(  k+n+m\right)  \beta}~L_{k+m+n}\ .
\end{align}
Here we have used (\ref{VirasoroWitt}) and an identity specific to the
realization at hand,%
\begin{equation}
L_{k}L_{m+n}=\left(  L_{0}+k\beta\right)  \left(  1+\frac{k\left(
1-\beta\right)  }{L_{0}+\left(  k+n+m\right)  \beta}\right)  L_{k+m+n}\ ,
\end{equation}
plus cyclic permutations of $k,m,n$. \ Note the simplification when $\beta=0 $
or when $\beta=1$. \ Similarly, we compute%
\begin{equation}
\left[  L_{p},L_{q},\left[  L_{k},L_{m},L_{n}\right]  \right]  =C~\left(
k-m\right)  \left(  m-n\right)  \left(  k-n\right)  \left(  p-q\right)
~\left(  1+\frac{\left(  k+m+n\right)  \left(  1-2\beta\right)  }%
{L_{0}+\left(  k+m+n+p+q\right)  \beta}\right)  L_{k+m+n+p+q}\ .
\end{equation}
The case $\beta=1/2$ stands out as particularly tasteful.%
\begin{equation}
\left.  \left[  L_{p},L_{q},\left[  L_{k},L_{m},L_{n}\right]  \right]
\right\vert _{\beta=1/2}=\frac{1}{4}\left(  k-m\right)  \left(  m-n\right)
\left(  k-n\right)  \left(  p-q\right)  L_{k+m+n+p+q}\ .
\end{equation}
This case allows us to quickly check the FI condition for the $L$s. \ We find
that it holds true.%
\begin{equation}
\left.  \left[  L_{p},L_{q},\left[  L_{k},L_{m},L_{n}\right]  \right]
-\left[  \left[  L_{p},L_{q},L_{k}\right]  ,L_{m},L_{n}\right]  -\left[
L_{k},\left[  L_{p},L_{q},L_{m}\right]  ,L_{n}\right]  -\left[  L_{k}%
,L_{m},\left[  L_{p},L_{q},L_{n}\right]  \right]  \right\vert _{\beta
=1/2}=0\ .
\end{equation}
Before checking the FI for other values of $\beta$, it is useful to first make
some general observations.

The 3-bracket of $L$s introduces a new operator given in the present
realization as a product. \ To check the FIs for the 3-algebra then requires
knowing how this new operator is affected under the action of a 3-bracket.
\ Since the 3-bracket is \emph{not} a derivation, in general, this requires
some new input, even when the new operator is realized as a product.
\ Therefore, we require several independent statements for the various
3-brackets. \ From the $su\left(  1,1\right)  $ enveloping algebra
expressions, we find\
\begin{align}
\left[  L_{k},L_{m},L_{n}\right]   &  =C~\left(  k-m\right)  \left(
m-n\right)  \left(  k-n\right)  ~M_{k+m+n}\ ,\nonumber\\
\left[  L_{p},L_{q},M_{k}\right]   &  =\left(  p-q\right)  ~\left(
L_{k+p+q}+\left(  1-2\beta\right)  k~M_{k+p+q}\right)  \ ,
\end{align}
where, in that realization,%
\begin{equation}
M_{k}=\frac{1}{L_{0}+k\beta}~L_{k}=\frac{\Gamma\left(  L_{0}+\beta\right)
}{\Gamma\left(  L_{0}+\beta+k\right)  }\left(  L_{1}\right)  ^{k}\ .
\end{equation}
In particular, $M_{0}=1$. \ The Casimir-dependent factors are tantamount to
operators, of course, and will eventually be absorbed into normalizations, but
we leave them explicit for now. \ To close the algebra, we also need to
consider two additional 3-brackets: $\left[  L_{p},M_{q},M_{k}\right]  $\ and
$\left[  M_{p},M_{q},M_{k}\right]  $.

So we compute some more in the enveloping algebra, to find%
\begin{equation}
\left[  M_{q},M_{k}\right]  =0\ ,\ \ \ \left[  L_{p},M_{q}\right]
=-qM_{p+q}\ , \label{L&MVirasoroWitt}%
\end{equation}
and then,%
\begin{align}
\left[  L_{p},M_{q},M_{k}\right]   &  =L_{p}\left[  M_{q},M_{k}\right]
+M_{k}\left[  L_{p},M_{q}\right]  -M_{q}\left[  L_{p},M_{k}\right] \nonumber\\
&  =-qM_{k}M_{p+q}+kM_{q}M_{p+k}\nonumber\\
&  =\left(  k-q\right)  M_{k+p+q}\ ,
\end{align}
upon using another identity valid in the $su\left(  1,1\right)  $ realization.
\ Namely,\footnote{An alternative presentation of the Lie algebra for the $L$s
and $M$s is given by taking generating elements $L_{0}$ and $M_{1}$, with the
condition $\left[  L_{0},M_{1}\right]  =-M_{1}$. \ Then $M_{n}\equiv\left(
M_{1}\right)  ^{n}$ and $L_{n}\equiv M_{n}\left(  L_{0}+n\left(
\beta-1\right)  \right)  $ leads to the algebra (\ref{VirasoroWitt}),
(\ref{L&MVirasoroWitt}). \ From this it is clear the $L$s depend on $\beta$,
but the $M$s do not.} \
\begin{equation}
M_{k}M_{p+q}=M_{k+p+q}\ .
\end{equation}
Thus we arrive at the remaining 3-brackets.%
\begin{equation}
\left[  M_{p},M_{q},M_{k}\right]  =0\ ,\ \ \ \left[  L_{p},M_{q},M_{k}\right]
=\left(  k-q\right)  ~M_{k+p+q}\ .
\end{equation}
The complete 3-algebra found through use of the $su\left(  1,1\right)  $
enveloping algebra is then%
\begin{gather}
\left[  L_{k},L_{m},L_{n}\right]  =C~\left(  k-m\right)  \left(  m-n\right)
\left(  k-n\right)  ~M_{k+m+n}\ ,\nonumber\\
\left[  L_{p},L_{q},M_{k}\right]  =\left(  p-q\right)  ~\left(  L_{k+p+q}%
+\left(  1-2\beta\right)  k~M_{k+p+q}\right)  \ ,\nonumber\\
\left[  L_{p},M_{q},M_{k}\right]  =\left(  k-q\right)  ~M_{k+p+q}%
\ ,\nonumber\\
\left[  M_{p},M_{q},M_{k}\right]  =0\ . \label{L&M}%
\end{gather}
In addition, the $su\left(  1,1\right)  $ realization gives the interesting
result that all higher N-brackets (i.e. totally antisymmetrized products of N
operators) are null. \ This follows from explicit calculation of the five
possible forms for 4-brackets, three of which exhibit nontrivial cancellations
of terms, to obtain%
\begin{equation}
0=\left[  L_{j},L_{k},L_{m},L_{n}\right]  =\left[  M_{j},L_{k},L_{m}%
,L_{n}\right]  =\left[  M_{j},M_{k},L_{m},L_{n}\right]  =\left[  M_{j}%
,M_{k},M_{m},L_{n}\right]  =\left[  M_{j},M_{k},M_{m},M_{n}\right]  \ .
\end{equation}
Consequently, all 5-brackets vanish upon being resolved into 4-brackets
\cite{C}, etc. Thus, all $(N\geq4)$-brackets vanish for the $su\left(
1,1\right)  $ realization of this infinite algebra. \ 

From the first two relations in (\ref{L&M}) it is now straightforward to check%
\begin{equation}
\left[  L_{p},L_{q},\left[  L_{k},L_{m},L_{n}\right]  \right]  =\left[
\left[  L_{p},L_{q},L_{k}\right]  ,L_{m},L_{n}\right]  +\left[  L_{k},\left[
L_{p},L_{q},L_{m}\right]  ,L_{n}\right]  +\left[  L_{k},L_{m},\left[
L_{p},L_{q},L_{n}\right]  \right]  \ , \label{LFI}%
\end{equation}
for any value of $C$ (and $\beta$). \ But of course, there are more possible
FIs involving both $L$s and $M$s, and these should also be checked. \ We
proceed to do this, after a bit of stream-lining. \ 

The overall Casimir factor in the first relation of (\ref{L&M}) is easily
eliminated by rescaling the charges by a fourth root of the Casimir, as in the
original $su\left(  2\right)  $ example of Nambu: \ $Q_{k}\equiv\frac
{1}{\sqrt[4]{C}}L_{k}$, $R_{k}\equiv\sqrt[4]{C}M_{k}$. \ However, the
remaining Casimir-dependent factor in the second relation of (\ref{L&M}) (i.e.
a central charge $z\equiv\left(  1-2\beta\right)  /\sqrt{C}$ after the
rescalings) is \emph{not} so easily removed. \ In any case we now abstract
from the $su\left(  1,1\right)  $ enveloping algebra results the following.

\section{Virasoro-Witt 3-Algebra}%

\begin{gather}
\left[  Q_{k},Q_{m},Q_{n}\right]  =\left(  k-m\right)  \left(  m-n\right)
\left(  k-n\right)  ~R_{k+m+n}\ ,\nonumber\\
\left[  Q_{p},Q_{q},R_{k}\right]  =\left(  p-q\right)  ~\left(  Q_{k+p+q}%
+z~k~R_{k+p+q}\right)  \ ,\nonumber\\
\left[  Q_{p},R_{q},R_{k}\right]  =\left(  k-q\right)  ~R_{k+p+q}%
\ ,\nonumber\\
\left[  R_{p},R_{q},R_{k}\right]  =0\ , \label{TernaryWitt}%
\end{gather}
where the central charge $z$ is effectively a parameter. \ For generic values
of $z$, by direct application of this ternary algebra, we now find that
3-brackets acting on 3-brackets satisfy the usual Leibniz-like rules (FIs)
\emph{except} when only one $R$ is involved. \ There are two such exceptions
out of twelve 3-on-3 possibilities. \ 

In an obvious notation, the twelve FI possibilities stem from each of the following:

\hspace{-0.25in}$%
\begin{array}
[c]{cccccc}%
\lbrack R,R,[R,R,R]]\ , & [R,R,[Q,R,R]]\ , & [Q,R,[R,R,R]]\ , &
[R,R,[Q,Q,R]]\ , & [Q,R,[Q,R,R]]\ , & [Q,Q,[R,R,R]]\ ,\\
\lbrack Q,Q,[Q,Q,Q]]\ , & [Q,Q,[R,Q,Q]]\ , & [R,Q,[Q,Q,Q]]\ , &
[Q,Q,[R,R,Q]]\ , & [R,Q,[R,Q,Q]]\ , & [R,R,[Q,Q,Q]]\ .
\end{array}
$

\noindent Ten of the FIs hold and behave as Filippov and Leibniz would
dictate. \ For example, for any $z$, the 3-algebra (\ref{TernaryWitt}) gives%
\begin{equation}
\left[  Q_{p},Q_{q},\left[  Q_{k},Q_{m},Q_{n}\right]  \right]  =\left[
\left[  Q_{p},Q_{q},Q_{k}\right]  ,Q_{m},Q_{n}\right]  +\left[  Q_{k},\left[
Q_{p},Q_{q},Q_{m}\right]  ,Q_{n}\right]  +\left[  Q_{k},Q_{m},\left[
Q_{p},Q_{q},Q_{n}\right]  \right]  \ ,
\end{equation}
as stated earlier in terms of $L$s. \ For another example,%
\begin{equation}
\left[  R_{p},R_{q},\left[  Q_{k},Q_{m},Q_{n}\right]  \right]  =\left[
\left[  R_{p},R_{q},Q_{k}\right]  ,Q_{m},Q_{n}\right]  +\left[  Q_{k},\left[
R_{p},R_{q},Q_{m}\right]  ,Q_{n}\right]  +\left[  Q_{k},Q_{m},\left[
R_{p},R_{q},Q_{n}\right]  \right]  \ .
\end{equation}
From (\ref{TernaryWitt}) the LHS trivially vanishes in this case, while the
three terms on the RHS cancel. \ 

The two exceptions, which for generic $z$ do \emph{not} obey FI
conditions,\ give instead
\begin{gather}
\left[  Q_{p},Q_{q},\left[  Q_{k},Q_{m},R_{n}\right]  \right]  -\left[
\left[  Q_{p},Q_{q},Q_{k}\right]  ,Q_{m},R_{n}\right]  -\left[  Q_{k},\left[
Q_{p},Q_{q},Q_{m}\right]  ,R_{n}\right]  -\left[  Q_{k},Q_{m},\left[
Q_{p},Q_{q},R_{n}\right]  \right] \nonumber\\
=\left(  4+z^{2}\right)  \left(  p-q\right)  \left(  k-m\right)  \left(
m-p-q+k\right)  n~R_{k+m+n+p+q}\ ,\nonumber\\
\nonumber\\
\left[  Q_{p},R_{q},\left[  Q_{k},Q_{m},Q_{n}\right]  \right]  -\left[
\left[  Q_{p},R_{q},Q_{k}\right]  ,Q_{m},Q_{n}\right]  -\left[  Q_{k},\left[
Q_{p},R_{q},Q_{m}\right]  ,Q_{n}\right]  -\left[  Q_{k},Q_{m},\left[
Q_{p},R_{q},Q_{n}\right]  \right] \nonumber\\
=\left(  4+z^{2}\right)  \left(  n-k\right)  \left(  k-m\right)  \left(
m-n\right)  q~R_{k+m+n+p+q}\ . \label{FailureIsAnOption}%
\end{gather}
Nevertheless, for the special cases $z=\pm2i$ the RHSs of
(\ref{FailureIsAnOption}) also vanish. \ Hence in these special cases all the
FIs hold for the algebra of (\ref{TernaryWitt}), making it a bona fide 3-algebra.

It is interesting that the special values $z=\pm2i$ are obtained in the
$su\left(  1,1\right)  $\ realization only in the \textquotedblleft
classical\textquotedblright\ limit of large Casimirs, $C=\beta\left(
1-\beta\right)  \rightarrow-\infty$, for which%
\begin{equation}
z^{2}=\frac{\left(  1-2\beta\right)  ^{2}}{\beta\left(  1-\beta\right)
}~_{\overrightarrow{\beta\rightarrow\pm\infty}}~-4\ .
\end{equation}
Perhaps this removes some of the mystery surrounding the FIs, which are
\emph{all} true statements for the proposed Virasoro-Witt 3-algebra for these
special values of $z$, since the FIs are known to hold for classical Nambu
3-brackets (as in (\ref{ClassicalNambu3BracketAlgebraFIs}) above). \ In this
context, we note the coefficient on the RHS of the first relation in
(\ref{TernaryWitt}) is given by $-a\cdot\left(  b\times c\right)  $ for three
vectors $a=\left(  1,k,k^{2}\right)  $, $b=\left(  1,m,m^{2}\right)  $, and
$c=\left(  1,n,n^{2}\right)  $.

As emphasized in \cite{C} there is nothing fundamental about the FI conditions
so far as associativity is concerned. \ Generally these \textquotedblleft
identities\textquotedblright\ fail to hold due to quantum corrections for
associative operator algebras $\hbar$-deformed from the Jacobian-like limit of
the classical Nambu brackets (CNBs) to become full-fledged quantum Nambu
brackets (QNBs). \ Although the FIs fail for (\ref{FailureIsAnOption}) for
generic $z$, there is an alternative identity following only from
associativity of the charge multiplication. \ This alternate identity must
hold if the associative algebra is consistent. \ The identity is
\begin{align}
&  \left[  \left[  A,B,C\right]  ,D,E\right]  +\left[  \left[  A,D,E\right]
,B,C\right]  +\left[  \left[  D,B,E\right]  ,A,C\right]  +\left[  \left[
D,E,C\right]  ,A,B\right] \nonumber\\
&  -\left[  \left[  D,B,C\right]  ,A,E\right]  -\left[  \left[  E,B,C\right]
,D,A\right]  -\left[  \left[  A,D,C\right]  ,B,E\right] \nonumber\\
&  -\left[  \left[  A,E,C\right]  ,D,B\right]  -\left[  \left[  A,B,D\right]
,C,E\right]  -\left[  \left[  A,B,E\right]  ,D,C\right] \nonumber\\
&  =3\left[  A,B,C,D,E\right]  \;.
\end{align}
This is the prototypical generalization of the Jacobi identity for odd QNBs
\cite{C}, and like the Jacobi identity, it is antisymmetric in all of its
elements. \ The RHS here is an \textquotedblleft
inhomogeneity\textquotedblright\ that illustrates a more general result:
\ \textit{The totally antisymmetrized action of odd }$n$\textit{\ QNBs on
other odd }$n$\textit{\ QNBs results in }$\left(  2n-1\right)  $%
\textit{-brackets.} \ For the case at hand, by explicit calculation using
(\ref{TernaryWitt}), we find for any value of $z$,%
\begin{align}
&  \left[  Q_{p},Q_{q},\left[  Q_{k},Q_{m},R_{n}\right]  \right]  +\left[
Q_{m},R_{n},\left[  Q_{k},Q_{p},Q_{q}\right]  \right]  +\left[  Q_{k}%
,R_{n},\left[  Q_{p},Q_{m},Q_{q}\right]  \right]  +\left[  Q_{k},Q_{m},\left[
Q_{p},Q_{q},R_{n}\right]  \right] \nonumber\\
&  -\left[  Q_{k},Q_{q},\left[  Q_{p},Q_{m},R_{n}\right]  \right]  -\left[
Q_{p},Q_{k},\left[  Q_{q},Q_{m},R_{n}\right]  \right]  -\left[  Q_{m}%
,Q_{q},\left[  Q_{k},Q_{p},R_{n}\right]  \right] \nonumber\\
&  -\left[  Q_{p},Q_{m},\left[  Q_{k},Q_{q},R_{n}\right]  \right]  -\left[
R_{n},Q_{q},\left[  Q_{k},Q_{m},Q_{p}\right]  \right]  -\left[  Q_{p}%
,R_{n},\left[  Q_{k},Q_{m},Q_{q}\right]  \right] \nonumber\\
&  =0\ .
\end{align}
Remarkably, there is \emph{no inhomogeneity} for this or any other 3-on-3
situation computed using (\ref{TernaryWitt}). Perhaps this is not too
surprising given that all $(N\geq4)$-brackets vanish for the $su\left(
1,1\right)  $ realization that led to (\ref{TernaryWitt}), although in any
other realization it would be necessary to specify all the 5-brackets, either
by direct calculation where possible, or by definition if necessary. Still, it
is reassuring that the proposed 3-algebra (\ref{TernaryWitt}) satisfies these
associativity-required consistency checks.

\section{Discussion}

Just as there are a countably infinite number of copies of $su\left(
1,1\right)  $ embedded in the centerless Virasoro-Witt algebra, so there are
an infinite number of BL algebras embedded in the (complexified) ternary
Virasoro-Witt algebra proposed here. \ At any given level, $L_{\pm n}/\sqrt
{n}$ and $L_{0}$ obey the $su\left(  1,1\right)  $ commutation relations
(\ref{su(1,1)}), with invariant $C_{n}=L_{+n}L_{-n}-\left(  L_{0}+n\right)
L_{0}$. \ The construction of the BL algebra at that level then proceeds as in
(\ref{NambuSU2}) and (\ref{NambuA4}) above, after complexifying the level's
$su\left(  1,1\right)  $ to obtain $su\left(  2\right)  $ in the well-known way.

It is a straightforward extension to include central charges in the Virasoro
algebra, as well as the $L$ and $M$ commutators,
\begin{equation}
\left[  L_{n},L_{m}\right]  =\left(  n-m\right)  L_{n+m}+cn^{3}\delta
_{n,-m}\ ,\ \ \ \left[  L_{n},M_{k}\right]  =-kM_{k+n}+bn^{2}\delta_{k,-n}\ ,
\end{equation}
where $b$ and $c$ are central, and to investigate their consequences for the
3-algebra. \ This will be discussed in detail elsewhere. \ Similar remarks
apply to the supersymmetric extension of (\ref{TernaryWitt}).

The astute reader will have noticed that (\ref{VirasoroWitt}) and
(\ref{L&MVirasoroWitt})\ form a sub-algebra of the Schr\"{o}dinger-Virasoro
algebra \cite{SV}. \ The extension of our 3-bracket results to include the
remaining charges of that larger algebra could be of interest.

Finally, we re-emphasize that our construction leading to (\ref{TernaryWitt})
used only 3-brackets defined as totally antisymmetrized operator products:
\ $\left[  A,B,C\right]  \equiv ABC-BAC+BCA-CBA+CAB-ACB$. \ Any other
definition of the 3-brackets might lead to a ternary algebra different from
that proposed here.

\paragraph{Acknowledgements}

This work was supported by NSF Award 0555603 and by the US Department of
Energy, Division of High Energy Physics, Contract DE-AC02-06CH11357.

\end{document}